\documentclass[]{agujournal2019}
\usepackage{apacite}
\usepackage{url} 
\usepackage[T1]{fontenc}
\usepackage{natbib}

\draftfalse

\journalname{Space Weather}

\begin{document}

%
%


\title{Statistical Analysis of the Correlation between Anomalies in the Czech Electric Power Grid and Geomagnetic Activity}

%
%




\authors{Tatiana V\'ybo\v{s}\v{t}okov\'a\affil{1}, Michal \v{S}vanda\affil{1,2}}
\affiliation{1}{Astronomical Institute, Charles University, V~Hole\v{s}ovi\v{c}k\'ach 2, 18000 Praha, Czech Republic}
\affiliation{2}{Astronomical Institute of the Czech Academy of Sciences, Fri\v{c}ova  298, 25165 Ond\v{r}ejov, Czech Republic}






\correspondingauthor{Michal \v{S}vanda}{michal@astronomie.cz}




\begin{keypoints}
\item We compare the series of disturbances recorded in the Czech electric power transmission network with the geomagnetic activity
\item The comparison is done in a statistical sense by considering only possible time scale of tens of days
\item We find indications that the mid-latitude power grid may be also affected by space weather events
\end{keypoints}

%
%


\begin{abstract}
Eruptive events on the Sun have an impact on the immediate surroundings of the Earth. Through induction of electric currents, they also affect Earth-bound structures such as the electric power transmission networks. Inspired by recent studies we investigate the correlation between the disturbances recorded in 12 years in the maintenance logs of the Czech electric-power distributors with the geomagnetic activity represented by the $K$ index.

We find that in case of the datasets recording the disturbances on power lines at the high and very high voltage levels and disturbances on electrical substations, there is a statistically significant increase of anomaly rates in the periods of tens of days around maxima of geomagnetic activity compared to the adjacent minima of activity. There are hints that the disturbances are more pronounced shortly after the maxima than shortly before the maxima of activity 

Our results provide indirect evidence that the geomagnetically induced currents may affect the occurrence rate of anomalies registered on power-grid equipment even in the mid-latitude country in the middle of Europe. A follow-up study that includes the modelling of geomagnetically induced currents is needed to confirm our findings. 

\end{abstract}

%
%

\section{Solar Activity}
The Sun consists of hot plasma interwoven with a magnetic field. These fields are created and amplified in the outer envelope of the solar body, rise and blend through the solar atmosphere. Since the outer envelope of the Sun is very dynamic (mainly due to convection), the magnetic field changes with time as well. The phenomenon associated with the existence and variability of the localised magnetic field is called \emph{solar activity} and consists of phenomena such as sunspots or prominences. Violent phenomena such as solar flares and coronal mass ejections (CMEs) may potentially have a dramatic impact on the Earth's environment, see e.g. a review by \cite{Gopalswamy2006} or recent studies by \cite{Krauss2015}, \cite{Goswami2018}, or \cite{Badruddin2019}. 

Solar flares are associated with a sudden reconnection of the magnetic field \citep[see a recent review by][]{shibata2011solar}. During the explosive reconnection, the coronal material is heated to tens of millions of degrees and becomes a source of X-rays. According to a standard (CSHKP) flare model, a disconnected plasma cloud (a plasmoid) wrapped in its own magnetic field is expelled into the interplanetary space in a form of a CME. There were also CMEs observed not obviously associated with a flare. Whether there indeed are two distinct classes of CMEs remains unclear \citep{Vrsnak2005}. Interplanetary CMEs have a great potential to strongly influence technologies on the Earth.

However, not only flares and consequent CMEs affect the close surroundings of the Earth. Moderate but principally similar effects may be caused by streams of the fast solar wind within the coronal holes or by the shocks resulting from the interaction of the fast and slow solar wind in the corotating interaction regions \citep[see e.g. a recent review by][]{Richardson2018}.

\section{Geomagnetic Activity and its Effects}
The interaction of the Earth's magnetic field with the solar wind shock wave or cloud of the magnetic field is very complex and cannot be described by a few sentences, see a review by \cite{Pulkkinen2007}. The interaction results in the disturbances of the Earth's magnetic field. These disturbances are in summary nicknamed \emph{geomagnetic activity}. Significant fluctuations of the geomagnetic activity are called geomagnetic storms. The causes, evolution and effects of the geomagnetic storms are comprehensively described elsewhere, e.g. in reviews by \cite{Tsurutani1997} or \cite{Lakhina2016}. 

The level of geomagnetic activity can be most easily expressed by measuring the Earth's magnetic field strength. From the measured geomagnetic field and its evolution in time a variety of indices of geomagnetic activity may be constructed. One of them is a $K$ index. It is a semi-logarithmic quantity describing changes in the amplitude of the horizontal component of Earth's magnetic field over a three-hour interval. When $K=0$, then the geomagnetic field is in a quiescent state, $K>5$ indicates a geomagnetic storm and $K=9$ indicates a superstorm. 

The interaction of the geomagnetic field with the variable solar wind induces changes in the terrestrial magnetic field at ground level, thereby possibly affecting the human infrastructure. The varying geomagnetic activity is connected with changes in the system of currents of the magnetosphere and ionosphere that generate a time-varying electric field  at the Earth's surface \citep[see e.g. a review by][and references therein]{Buonsanto1999}. This geoelectric field, in turn, gives rise to geomagnetically induced currents (GIC) in the conductive structures on the Earth's surface and also below to a substantial depth \citep[see a review by][]{Pulkkinen2017}.

The GICs run through conducting regions of the Earth and seas and also between the endings of grounded conductors. They can produce damage in the system attached to the conductor such as a signalling network associated with railways \citep{eroshenko2010effects}, pipelines \citep{pulkkinen2001recordings} and particularly in power networks \citep{pirjola2000geomagnetically}. 

The presence of GICs in the electric power transmission network may interfere with their normal operation and cause damage resulting in a failure or service disruption. GICs constitute a possible trouble for network transformers \citep[e.g.][]{Kappenman2007,Molinski2002}. The quasi-direct GICs lead to a half-cycle saturation by shifting the ($B$-$H$) curve to a strongly non-linear regime with a large portion of the magnetising current. In this regime harmonics are generated, heavy reactive power appears, the voltage may drop and even blackout may take place. Under half-wave saturation, the magnetic flux is leaked from the transformer leading to local heating. Oil in the cooling bath degrades, gassing appears. In the extreme event, the transformer core may start melting. The electrical current in the network changes its waveform to a distorted sinusoidal. Protective relays may react to the peak value of the corrupted waveform, evaluate it as overvoltage or overcurrent and trip the equipment. Last but not least, the presence of GIC impacts the stability of the system frequency with possibly damaging effects on generating stations. 

Most of the available work deals with the immediate effects registered on the infrastructure during or shortly after significant events in solar activity. A large portion of studies was published on  Qu\'ebec blackout in 1989, a very nice summary can be found in a report of \citet{NERC} or by \citet{kappenman1997} and later by \cite{bolduc2002}. Most of the literature deals with high-latitude locations because the effects of space-weather events are the strongest there. Nonetheless, the effects were observed during the storms and sub-storms also in mid- and low-latitudes, such as South Africa \citep{gaunt2007}, Japan \citep{Watari2009}, New Zealand \citep{Marshall2012}, Spain \citep{Torta2012}, or Italy \citep{Tozzi2019}. Large GICs in several mid- and low-latitude regions were also registered \citep{Kappenman2003}.

Only recently, research has been focused on the impact of fluctuating solar activity on the network infrastructures over an extended period. \citet{schrijver2013disturbances} studied the disturbances in the US electric power transmission network for the period from 1992 to 2010. They found, with more than 3$\sigma$ significance, that approximately 4\% of the disturbances in the US electric power transmission network are attributable to strong geomagnetic activity and associated GICs. This study was followed by \citet{schrijver2014assessing}, where the insurance claims due to the disturbances in the electric power transmission network were studied. The 4\% attribution to solar activity effects was confirmed. Moreover, in upper 5\% stormiest days the number of insurance claims increased by 20\%. In the upper third stormiest days, the total increase of the insurance claims was about 10\% larger compared to the quiescent period. 

The aim of this work is to make comparable analyses for disturbances recorded in the Czech electric power transmission network and determine the relationship between the anomalies of the grid components and increased geomagnetic activity in the Czech Republic.

To our knowledge, a single study was published dealing with GICs in the Czech Republic. \citet{hejda2005} analysed the pipe-to-soil voltages measured in oil pipelines in the Czech Republic during the Halloween storms in 2003. They showed that the simplest plane-wave and uniform-Earth model gives results that correspond well to the measured pipe-to-soil voltages. To complete the picture, the study performed in a neighbouring country must be noted. \citet{bailey2017} modelled and measured the GICs in the Austrian electric power grid. They demonstrated that the Austrian electric power transmission network is susceptible to large GICs in the range of tens of amperes, particularly from strong geomagnetic variations in the east--west direction. That is due to the low surface conductivity in the region of the Alps. 

\section{Data}
Although disturbances in power grids associated with strong solar activity events are known in Europe, no one has done a statistical study for European countries covering an extended period. There may be several reasons: for example, to carry out a large-scale study for a territory comparable to the United States is extremely difficult. Even though the power networks of the European countries are interconnected, each European state has its own national operator and many local distributors, which deal with disturbances according to their own internal regulations and follow different procedures. To combine different datasets may thus be extremely difficult. 

\subsection{The Electric Power Transmission Network in the Czech Republic}
The Czech Republic lies in central Europe and is extended in the east--west direction (about~500 km length) compared to the ``width'' in the south--north direction (about 280~km). In terms of the electric power grid, the spine of the power network is operated by the national operator \v{C}EPS, a.s., which maintains the very-high-voltage (400~kV and 220~kV) transmission network, and connects the Czech Republic with the neighbouring countries. \v{C}EPS also maintains the key transformers and electrical substations in the transmission network. The area of the state is then split into three regions, where the electricity distribution is under the responsibility of the distribution operators. The southern part is maintained by E.ON Distribuce, a.s., the northern part by \v{C}EZ Distribuce, a.s., and the capital city of Prague is maintained by PREdistribuce, a.s. All three distributors maintain not only very-high-voltage (110 kV) and high-voltage (22 kV) power lines, but also connect the consumers via the low-voltage (400 V) electric power transmission network. 

\subsection{Logs of Anomalies}
After years of delicate negotiations, we managed to obtain the maintenance logs from all the operators mentioned in the previous section. We obtained essentially the lists of disturbances recorded in the maintenance logs by the company technicians with their dates and many more details, which included also the probable cause of the anomaly. The lists contained not only the events of the equipment failure (e.g. defects), but also the events on the power lines, such as the repeated unplanned switching, power cuts, or service anomalies. By mutual non-disclosure agreement with the data providers, the datasets were anonymised and must be presented as such. The total time span is 12 years, but the span of individual maintenance logs provided by the operators is shorter, varying between 6 to 10 years. 

The inhomogeneous datasets were split into twelve subsets D1--D12, which were investigated separately. Each sub-dataset was selected so that it contained only events occurring on devices of a similar type and/or with the same voltage level and were recorded by the same operating company. 

We first went through an extensive manual check of the obtained logs, when we excluded the events which principally could not be related to the geomagnetic activity. That is we excluded the defects that occurred prior to putting the equipment into operation (i.e. manufacturing defects that were revealed during the testing period) or anomalies caused by other, space-weather unrelated effects. We downselected either erroneous records, the records connected with other large-scale natural disasters (floods) and records where the human factor was a prime cause for the anomaly (traffic accidents, customer's error). The fraction of the downselected events is given in the last column of Table~\ref{tab:datasets}. We did not exclude any events for datasets D7--D12. The fractions are larger for D3 and D4, where the large-area floods played a large role and the distributor recorded also planned upgrades as anomalies in the logs.  In datasets D5 and D6 a large fraction of records was erroneous (the date was not specified). In the same datasets, we excluded also power-line breaches caused by accidents during construction works from further analysis. 

\begin{figure}
    \centering
    \includegraphics[width=0.9\textwidth]{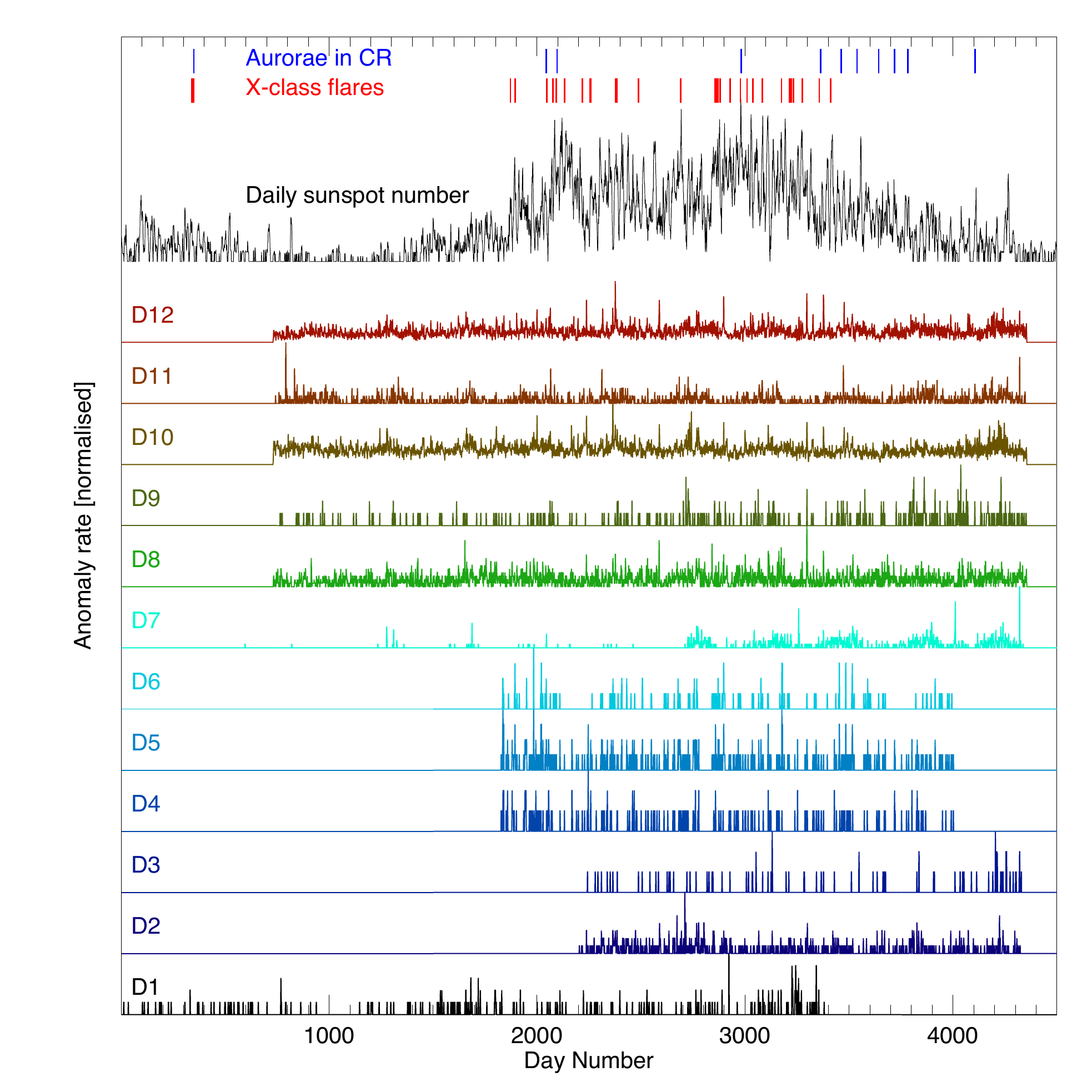}
    \caption{A representation of anomaly rates as registered in the various datasets in time. At the top of the figure we also plot the relative sunspot number for reference and indicate the dates, when X-class flares ignited and when aurorae were seen in the Czech Republic.}
    \label{fig:datasets}
\end{figure}

The dataset descriptions are summarised in Table~\ref{tab:datasets} and visualised in Fig.~\ref{fig:datasets} (where individual datasets were normalised to their corresponding maximum for displaying purposes due to the large spread of typical anomaly rates across datasets). We aim to study only the anomaly rates (an daily series of the counts of anomalies occuring in the same day) so that from the logs we kept only the date on which the event occurred and did not consider any other detail. This reduction was done for two reasons. First, the number of events was quite low in most cases (a few hundred per year usually) and further splitting would lower a statistical significance. Second, the records in the log were quite inhomogeneous even within the same log, because the forms entering the database were filled by different persons. The final clean datasets are one-by-one compared with the level of geomagnetic activity using statistical methods. 

For the sake of completeness, we merged all the subsets in one series nicknamed DX, which we analysed using the same methodology as the subsets. 

\begin{table}[ht]
    \caption{Datasets analysed in this study}
    \label{tab:datasets}
    \centering
    \begin{tabular}{l|llll}
      {\bf Dataset} & {\bf Voltage level} & {\bf Type} & {\bf Span} & {\bf Fraction}\\
      {\bf ID} & {\bf } & {\bf } & {\bf } & {\bf removed}\\
      \hline
      D1 & very high voltage & equipment: transformers, & 9 years\\
         &                   & electrical substations&  & 4\%\\
      D2 & high voltage & equipment & 6 years & 0.2\% \\
      D3 & very high voltage & equipment & 6 years & 10\% \\
      D4 & high and low voltage & power lines & 7 years & 13\%\\
      D5 & high and low voltage & equipment and power lines & 7 years & 58\%\\
      D6 & high and low voltage & equipment & 7 years & 55\% \\
      D7 & very high voltage & power lines & 10 years & 0\% \\
      D8 & high voltage & transformers & 10 years & 0\% \\
      D9 & very high voltage & transformers& 10 years & 0\% \\
      D10 & very high and high voltage & electrical substations& 10 years& 0\% \\
      D11 & very high voltage & power lines& 10 years& 0\%  \\
      D12 & high voltage & power lines & 10 years & 0\% \\
      \hline
      DX & \multicolumn{4}{l}{\ \ \it joined series of all the above}
    \end{tabular}
\end{table}

\subsection{Geomagnetic Activity}
The selection of an appropriate index to assess the effects of solar/geomagnetic activity to power grids is a delicate issue \citep[see the discussion e.g. in][]{schrijver2013disturbances}. We realised that none of the solar indices is suitable because events on the Sun (flares, CMEs) may have a different geoeffectivity. Still, the sunspots number or occurrence of X-class flares may serve as a secondary index to discuss various effects in the Sun-Earth connections. 

For the purpose of this study, we thus relied on the measurements of the geomagnetic field. We used the data from the nearest measuring station, the Geomagnetic Observatory Budkov in \v{S}umava mountains, operated by the Geophysical Institute of the Czech Academy of Sciences. They produce minute-by-minute measurements of the geomagnetic field vector. The measurements of the geomagnetic field were downloaded from the Intermagnet data archive, the gaps in the measurements (only 180 minutes over more than 13 years) were filled by using the measurements from Chambon-la-For\^{e}t station in France to have an uninterrupted data series.  

From these measurements, we constructed a $K$ index, which is typical for characterising the level of geomagnetic activity in similar applications. The $K$ index was calculated in using an FMI \citep{Sucksdorff1991} method, with a limit of 500~nT for $K=9$. To obtain daily values that could be compared with the anomaly rate with daily granularity we averaged 3-hour $K$ indices in each day. We realise that it may not be wise to average the semi-logarithmic quantity as the averaged values will no longer hold the physical meaning of the original quantity. For our study, we are not using the absolute values of $K$ index as a reference for the level of geomagnetic activity. We seek for local minima and maxima of geomagnetic activity only. For such purpose the averaged $K$-index series is a suitable quantity, despite possible interpretation issues.   

\section{Methods}
The goal of our work is to compare both data series of a different kind using various statistical tests. The correlation coefficient between the $K$ index and the anomaly rates of all series computed over the entire interval is practically zero, which is in agreement with a study by \citet{schrijver2013disturbances} principally similar to ours. The reason probably is that not all the power-grid disturbances are expected to occur immediately after the exposure to the increased geomagnetic activity. For a stringent example we need to note that in the case of the failure of the step-up transformer at Salem~2 generator station in 1989, the damage was discovered during a routine test a week after the exposure to large-amplitude GICs \citep{NERC}. Yet, the device was written off. \cite{Koen2003} point out the cumulative effects of GICs in the transformers. The dissolved-gas analysis records in the South-African transformers indicated that the deterioration continues after the initial damage caused by GICs. The time of the collapse is influenced by the transformer loading and possibly also by other stresses. Thus there may be a time delay with an unknown value, which depends on many conditions in the activity and on the device itself. E.g. for the Greek power grid, \citet{zois2013solar} found the delay to be up to several years. 

We were forced to use a different approach than a computation of the correlation coefficient. Our methodology is consistent with retrospective cohort study with tightly matched controls. First, we used the binomial test to check the hypothesis that around the maxima of geomagnetic activity the increase of the anomaly rates occurs, which is compared to the anomaly rate in a nearby minimum of activity. To do so, we searched for local minima and maxima in the $K$-index series. Around each minimum and maximum, we drew an interval with a total length of $W$ days. The window was positioned such that its centre was placed at the local minimum. For the maximum period, the window was placed \emph{after} the peak of activity, beginning at the local maximum. The maxima and minima were paired together so that they did not overlap and that the corresponding pairs were close in time. By such a selection we attempt to avoid long-term or secular trends in the logs of anomalies caused by the evolution of the grid in time and also seasonal changes due to the weather and loading. We performed our tests for a selection of values of $W$, these intervals served as accumulation windows for the series of power-grid disturbances. For each $W$ we had $n_{\rm i}$ pairs of maxima and minima of geomagnetic activity. 

In the selected intervals we counted the total number of anomalies $N_{\rm h}$ falling into the maximum intervals (that is during increased geomagnetic activity) and the number of anomalies $N_{\rm l}$ falling into the minimum intervals (that is in the low-activity periods). We only remind that these numbers depend on the length of $W$ of the accumulation window. 

In the case when increased geomagnetic activity on average induces a subsequently larger anomaly rate of power-grid devices, we would expect the relation
\begin{equation}
    N_{\rm h} > N_{\rm l}.
    \label{eq:inequality}
\end{equation}

Even when the above inequality holds, its statistical significance must be tested, for which we use the binomial test. The binomial test states the probability $P$ that the registered differences between $N_{\rm h}$ and $N_{\rm l}$ are in accordance with the model. Our model is the reversed hypothesis, that says there is no difference between the number of anomalies registered in the periods around local maxima of activity and local minima of activity. If $P$ is lower than 5\% (our selection of statistical significance), then we reject the reversed hypothesis. In such a case, we obtained an indication that indeed, there is a statistically significant increase in anomaly rates after the maximum of the geomagnetic activity. $P$ is computed as

\begin{equation}
P_{\rm h,l}=2\sum_{k=x}^{n} {{n}\choose{k}} p^k(1-p)^{n-k}
\label{eq:probability}    
\end{equation}
where for testing the pair $N_{\rm h}$ and $N_{\rm l}$, $n=N_{\rm h}+N_{\rm l}$ denotes the total number of anomalies in two sets of chosen intervals. The parameter $p$ states the model-expected probability of the disturbance occurring during the high-activity intervals. In the tested (that is in the reversed) hypothesis we assume that the probability of the disturbance occurring during the maximum or the minimum be the same, i.e. $p=1/2$. Finally, $x=\max(N_{\rm h},N_{\rm l}$). 

The binomial test is a principal approach in our study to assess the possibility of the disturbance rate to be affected by the geomagnetic activity. It gives us the answer to the question whether it is possible, for the given dataset and the accumulation window $W$, to register a difference in the anomaly rates occurring during the minima and maxima of the geomagnetic activity. It gives us a qualitative answer. 

To quantify the difference in the anomaly rates for both different situations we evaluated a relative risk $R$. It is a common quantity in e.g. cohort epidemiological studies or when testing the effectivity of the vaccination. It uses two samples: the sample which was exposed to a certain causal attribute (those are the testing object with the vaccine) and a control sample which was not exposed. Then the number of positive and negative cases in both samples are compared. The numbers of disturbances in the intervals of an increased and decreased geomagnetic activity are not suitable quantities for the calculation of the relative risk, because the cohort-study approach uses a binary ``yes/no'' flag in the description of the individuals in the exposed and control samples. Thus we constructed a different statistical series. For both the intervals around the local maxima of the geomagnetic activity (the sample with the causal attribute) we computed the number of days in which disturbances were registered and the number of days without disturbances. The analogous two numbers were computed for the interval around the local minima (the control sample without the causal attribute). Then we computed the relative risk as
\begin{equation} 
R=\frac{a}{a+b}/{\frac{c}{c+d}},
\label{eq:risk}
\end{equation}
where $a$ is the number of days with anomalies and $b$ without them, both for intervals with increased solar activity. For intervals with lower solar activity, $c$ is the number of days with an anomaly and $d$ is the number of days without any. 

The relative risk is 1 if there is no difference between the two groups differing in the causal attribute. If $R<1$ then more often positive cases occur in a group without a causal attribute (i.e. contrary to expectation), if $R>1$ then positive cases occur more often in a causal attribute group. 

The two tests described above may give us an indication of a statistically significant increase in the anomaly rates in the periods of the increased geomagnetic activity. These tests still do not prove the causal link, the binomial test is an ``advanced correlation measure'' to some extent. If the increased anomaly rates are indeed caused by the increased geomagnetic activity, where a positive but unknown time lag may play a role, one would expect that the anomaly rates are larger after the geomagnetic activity maximum than before. Thus we compared the number of disturbances in the intervals of length $W$ immediately before the local maximum with the number of disturbances in the intervals of the same length placed immediately after the maximum. We compared the mean daily anomaly rates in the two intervals (we would expect the mean to increase after the local maximum), the relative mean increase in the units of standard deviation of the daily anomaly rates, and of course we ran a binomial test evaluated by (\ref{eq:probability}) to test the statistical significance. 

A similar comparison was done for the minima. For the minima, we either expect the statistical parameters not changing much before and after the minimum. In the case when the descend to the local minimum was very steep and close to some much larger maximum, we could anticipate the statistics before the minimum be affected by the previously larger geomagnetic activity and then the mean anomaly rates before the minimum should be larger than after the minimum. 


\section{Results}
To perform the statistical analysis outlined in the previous section we wrote a series of programs in {\sc Python} that allowed us to study the effects of various parameters in the code on the results. Such an approach also ensured that the processing of all the data files was done in exactly the same way. The program had two principal inputs for each run: the series of daily-averaged $K$ index and the series of dates when anomalies on the electric power transmission network occurred in the given dataset. The analysis was executed for a set of window lengths $W$. 

\begin{figure}
    \centering
    \includegraphics[width=0.7\textwidth]{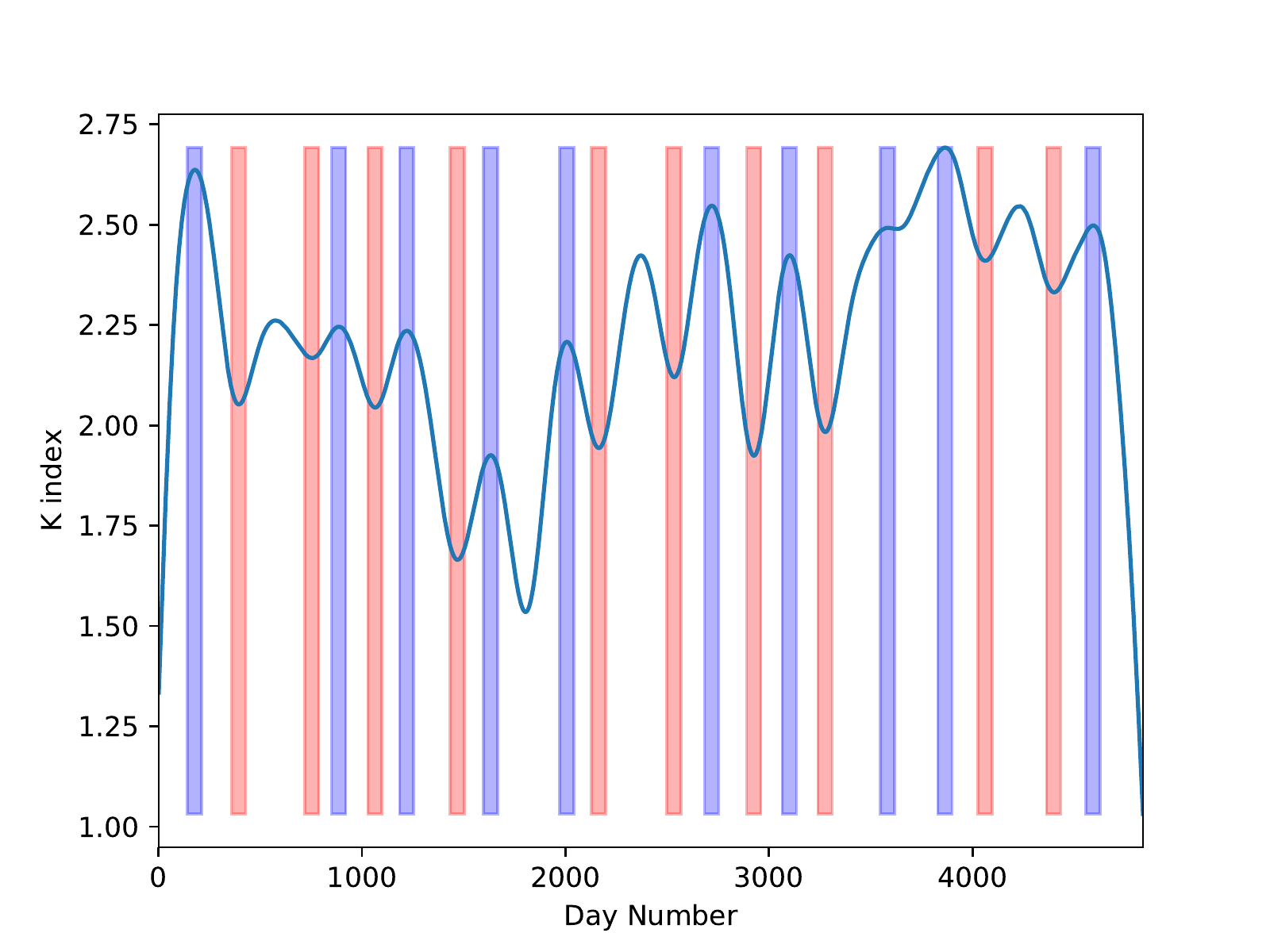}
    \caption{Daily averaged $K$ index (blue solid curve) where local minima and maxima were detected by our code. Around these extrema the intervals with a width of $W$~days are indicated by bars. The code searches for non-overlapping pairs of maximum (blue) and minimum (red) that are closest in time. This particular plot is for $W=70$~days.}
    \label{fig:intervals50}
\end{figure}

The code first smoothed the series of $K$ indices with a boxcar window with a width of $W$, so that the time-scales of the two series to be compared were similar. In the smoothed series the code then detected local maxima and minima from the first derivative. Then the neighbouring minima and maxima were paired together so that their paring minimises the possible effects of the secular development of the electric power transmission network that could possibly lead to the trends in the anomaly rates. On the other hand, the requirement was that the neighbouring minima and maxima do not overlap within $W$ bounds. In such a case the next neighbouring minimum was selected for that maximum. We also ensured that the maximum+minimum pairs were unique. An example of the local extrema detection and their pairing is given in Fig.~\ref{fig:intervals50}, a complete set is then given in Figs.~S1--S11 in the Supporting Information. 

The code then counted the number of disturbances falling into the maxima and minima and evaluated the probability given by (\ref{eq:probability}) and computed the risk given by (\ref{eq:risk}). For each maximum and each minimum, the code then evaluated the number of anomalies falling into the intervals of width $W$ occurring immediately before the maximum (or minimum) and after it. 

We found that the results depend strongly on the selection of $W$. Smaller windows increase the noise levels because lesser numbers of disturbances fall into the tested intervals. Also, shallower local minima and/or maxima are present in the series of $K$ index due to the smaller smoothing. On the other hand, too large values of $W$ cause the $K$ index series to be overly smoothed and the pairing of the neighbouring minima and maxima gets difficult because the requirement of the non-overlapping intervals is too strong. As a consequence, the ``neighbouring'' maximum+minimum pairs may be hundreds of days apart. The interpretation of such comparison is then complicated as over a period of hundreds of days the secular trends connected with the network development or seasonal changes increase their importance. In the following, we thus discuss mainly the results obtained for windows $W=(30,50,70)$~days.

We found that in most cases the differences in the number of anomalies in the maxima and minima are not statistically significant. That is also demonstrated in an example Table~\ref{tab:statistics70}. A complete set of tables for all investigated values of $W$ are given in the Tables S12--S22 in the Supporting Information. 

\begin{sidewaystable}
\caption{Statistical analysis of disturbances in the Czech electric power transmission network for the 70-day window. For datasets D1--D12 and DX we give the number of interval pairs, the total number of reported disturbances in the periods of increased activity and decreased activity. Then we give the probabilities $P$ with which the differences in the number of anomalies between two intervals are due to chance. In the last section we give necessary values for the computation of the relative risk $R$ given by (\ref{eq:risk}), and also the values of $I_{\rm r}$ given by (\ref{eq:Ir}).}
\label{tab:statistics70}
    \begin{tabular}{l|llll|llllll}
Dataset ID & Intervals & $N_{\rm h}$ & $N_{\rm l}$ & $P_{\rm h,l}$  & $a$ & $b$ & $c$ & $d$ & $R$ & $I_{\rm r}$ \\ 
\hline 
D1 & 7 & 46 & 22 & 0.0049 & 46 & 444 & 22 & 468 & 2.09091 & 0.07\\
D2 & 4 & 74 & 62 & 0.34559 & 74 & 206 & 62 & 218 & 1.19355 & --\\
D3 & 4 & 8 & 13 & 0.38331 & 8 & 272 & 13 & 267 & 0.61538 & --\\
D4 & 4 & 13 & 22 & 0.17547 & 13 & 267 & 22 & 258 & 0.59091 & --\\
D5 & 4 & 32 & 33 & 1.0 & 32 & 248 & 33 & 247 & 0.9697 & --\\
D6 & 4 & 19 & 12 & 0.28104 & 19 & 261 & 12 & 268 & 1.58333 & --\\
D7 & 8 & 84 & 31 & $<10^{-5}$ & 84 & 476 & 31 & 529 & 2.70968 & 0.53\\
D8 & 7 & 541 & 484 & 0.08022 & 298 & 192 & 279 & 211 & 1.0681 & --\\
D9 & 7 & 46 & 57 & 0.32448 & 44 & 446 & 51 & 439 & 0.86275 & --\\
D10 & 7 & 8661 & 7317 & $<10^{-5}$ & 490 & 0 & 490 & 0 & 1.0 & 0.13\\
D11 & 7 & 285 & 120 & $<10^{-5}$ & 194 & 296 & 92 & 398 & 2.1087 & 0.74\\
D12 & 7 & 7073 & 5543 & $<10^{-5}$ & 489 & 1 & 489 & 1 & 1.0 & 0.22\\
\hline
DX & 9 & 19462 & 13806 & $<10^{-5}$ & 566 & 64 & 497 & 133 & 1.13883 & 0.37\\
    \end{tabular}
\end{sidewaystable}

Only the datasets D1, D7, D10, D11, and D12 seem to indicate a statistically significant increase of $N_{\rm h}$ coherent with the working hypothesis. These five datasets behave the same for all windows $W$ in the range of 30 to 120 days. We note that D1 aggregates disturbances on the equipment of the spinal transmission network. Datasets D7, D10, D11, and D12 record disturbances on the high-voltage and very-high-voltage power lines and also on electrical substations. The joined series DX also indicates a significant increase of anomaly rate in the periods around local maxima of geomagnetic activity compared to neighbouring minima. Differences registered in the remaining datasets are not statistically significant. 

\begin{sidewaystable}
    \caption{Comparison of the anomaly rates around the local minima and maxima. In the first section we give mean values ($\mu$) and standard deviations ($\sigma$) for the intervals of length $W=70$~days before local maxima (subscript ${-\rm max}$) and after the maxima (subscript ${\rm max}+$). In the second section analogous parameters are evaluated for local minima. In the third section we give the relative increase of the means in the units of the standard deviation evaluated before the extrema. In the last section we give the probability that the differences in the interval before the maximum and after the maximum are due to chance.}
    \label{tab:minmax70}
\begin{tabular}{l|llll|llll|ll|l}
Dataset ID & $\mu_{-{\rm max}}$ & $\mu_{{\rm max}+}$ & $\sigma_{-{\rm max}}$ & $\sigma_{{\rm max}+}$ &  $\mu_{-{\rm min}}$ & $\mu_{{\rm min}+}$ & $\sigma_{-{\rm min}}$ & $\sigma_{{\rm min}+}$ & $\frac{I_{\rm max}}{\sigma_{-{\rm max}}}$ & $\frac{I_{\rm min}}{\sigma_{-{\rm min}}}$ & $p_{{+}{-}}$  \\ 
\hline 
D1 & 0.44 & 0.66 & 0.62 & 0.79 & 0.34 & 0.29 & 0.61 & 0.51 & 0.343 & -0.094 & 0.03355\\
D2 & 1.06 & 1.06 & 0.77 & 0.79 & 0.77 & 0.71 & 0.72 & 0.83 & 0.0 & -0.079 & 0.42219\\
D3 & 0.11 & 0.11 & 0.32 & 0.32 & 0.16 & 0.11 & 0.36 & 0.32 & 0.0 & -0.118 & 0.75391\\
D4 & 0.49 & 0.19 & 0.58 & 0.42 & 0.29 & 0.33 & 0.54 & 0.53 & -0.518 & 0.08 & 0.34493\\
D5 & 0.79 & 0.46 & 0.69 & 0.65 & 0.59 & 0.43 & 0.71 & 0.6 & -0.473 & -0.222 & 1.0\\
D6 & 0.33 & 0.27 & 0.47 & 0.53 & 0.34 & 0.11 & 0.58 & 0.32 & -0.122 & -0.392 & 0.15159\\
D7 & 1.1 & 1.2 & 0.93 & 0.82 & 0.43 & 0.39 & 0.55 & 0.54 & 0.108 & -0.078 & 0.51041\\
D8 & 6.6 & 7.73 & 2.85 & 3.62 & 7.61 & 6.49 & 3.5 & 2.56 & 0.396 & -0.323 & 0.00002\\
D9 & 0.71 & 0.66 & 0.94 & 0.89 & 0.7 & 0.6 & 0.82 & 0.85 & -0.061 & -0.122 & 0.00109\\
D10 & 112.99 & 123.73 & 21.08 & 21.2 & 101.09 & 103.49 & 19.34 & 20.01 & 0.51 & 0.124 & 0.00073\\
D11 & 4.29 & 4.07 & 3.06 & 2.23 & 1.44 & 2.13 & 1.31 & 1.6 & -0.07 & 0.525 & 0.81275\\
D12 & 87.87 & 101.04 & 19.02 & 20.4 & 82.93 & 76.57 & 20.31 & 20.11 & 0.692 & -0.313 & $<10^{-5}$\\
\hline
DX & 256.97 & 278.03 & 46.59 & 45.49 & 196.86 & 206.6 & 38.35 & 40.08 & 0.452 & 0.254 & $<10^{-5}$
\end{tabular}

\end{sidewaystable}

The relative risk as we defined it turned out to be not very useful in the analysis we performed. The $R$ value is meaningless when the number of recorded disturbances is low due to the statistical insignificance of $R$. On the other hand, when the number of disturbances is large, as in e.g. D12, then $R=1$ by definition. Due to the daily granularity of the anomaly logs, in such a case there always is at least one disturbance each day in which the data exist. Then obviously the $R$ is meaningless again. Thus it is useful only in the case when the total number of registered disturbance events is between say 100 and 300. We cannot increase the granularity in time of the populated datasets, thus we cannot choose another suitable parameter to be tested to compute the relative risk. One option would be to split the populated datasets in more datasets e.g. grouped by the geographical location or another criterion. We did not proceed in this way because we do not have confidence in a gain of such a non-trivial manual sorting of the inputs. 

Instead, we used a different method to quantify the increase of disturbances in the maxima of geomagnetic activity. For each dataset with positive indication of the increase ($P_{\rm h,l}<0.05$) we searched for the largest $N_{\rm h1}$ while keeping $N_{\rm l}$ constant, $N_{\rm h1} \geq N_{\rm l}$, for which $P_{\rm h1,l} \leq 0.01$. This would be the boundary value, where we could not reject the hypothesis that the differences are due to chance with a 1\% significance level. Note that for this task we chose a stronger limit of the statistical significance level. The relative increase with a statistical significance would be defined as 
\begin{equation}
    I_{\rm r}=(N_{\rm h}-N_{\rm h1})/N_{\rm h1}.
    \label{eq:Ir}
\end{equation}
The values of $I_{\rm r}$ are also given in Table~\ref{tab:statistics70} and the supplementary tables S12--S22 in the Supporting Information. 
In the case of D1, D7, D10, D11, and D12 $I_{\rm r}$ is between 7 and 70 per cent. 


We remind that the intervals of length $W$ were always centred on the local minimum, whereas they were placed \emph{after} the local maximum. The motivation was that we expect some delay in the occurrence of the disturbance after the exposure to larger GICs. To test how important this selection is for the results we ran our codes again when centering the intervals on the dates of maxima of geomagnetic activity. The results did not change significantly. 

On the other hand, it is worth testing whether the anomaly rate increases after the maximum of geomagnetic activity. Such a test was achieved by comparing the number of anomalies and their statistical properties in a window of length $W$ before the maximum with the window of the same length after the maximum. As for the considered statistical properties, we evaluate the mean daily anomaly rate for each dataset individually, computed over the window $W$-days long before the local maxima (indicated as $\mu_{\rm -max}$) and its variance (indicated as $\sigma_{\rm -max}$). Same statistical indicators were computed for $W$-days long window after the local maxima (these are referred to as $\mu_{\rm max+}$ and $\sigma_{\rm max+}$). The increase of the daily means $I_{\rm max}$ is given by
\begin{equation}
    I_{\rm max}=\mu_{\rm max+}-\mu_{\rm -max}.
\end{equation}

As a comparison, we derived the same parameters for the local minima (indicated by \emph{min} in their lower indices). The results are compatible with our hypothesis only if there is an increase in the mean daily anomaly rate after the local maxima and there is more-or-less no change or a decrease around the local minima. Example results again for $W=70$~days are shown in Table~\ref{tab:minmax70}. A complete set of tables for all investigated values of $W$ are given in the Tables S23--S33 in the Supporting Information. 

The statistical significance of the recorded differences before and after local maxima of geomagnetic activity is low for most of the datasets. The differences with larger statistical significance are present in case of datasets D1, D10, and D12, where also the significant difference between $N_{\rm h}$ and $N_{\rm l}$ was found. In those datasets, we observe an average increase of the anomaly rates by (0.3--0.7)$\sigma$ (about 10-30\% of the daily means) after the maxima, whereas in the control series around the minima we see either a decrease or a much lower increase than in case of the maxima. Furthermore, we record a statistically significant increase around the maxima in dataset D8. This dataset also recorded $N_{\rm h} > N_{\rm l}$ (see Table~\ref{tab:statistics70}), but the computed probability $P_{\rm h,l}=0.08$ was larger than the chosen 5\% threshold (the dataset is only 7 anomalies away from the chosen statistical significance threshold). D8 records the disturbances occurring on high-voltage transformers. 

\section{Concluding Remarks}
We searched for a ``correlation'' between the occurrence of disturbances on the electric power transmission network in the Czech Republic and the geomagnetic activity described by the $K$ index. The logs of disturbances were split into twelve different datasets according to the data provider (the company), voltage level, and a class of equipment. For the sake of completeness, we also investigated the merged dataset using the same methodology. 

We compared the anomaly rates in the periods tens of days long around local maxima of geomagnetic activity with the periods around local minima of geomagnetic activity using statistical methods. The statistical significance of this comparison highly depends on the size of the sample. For sparse datasets the results are inconclusive, so one cannot say whether there are differences in recorded rates of power-network anomalies in the periods exposed to a local maximum of geomagnetic activity as compared to a nearby local minimum of activity.  

For the populated datasets which record mostly the disturbances on power lines and electrical substations on the high-voltage and very-high-voltage levels, we see a statistically significant increase of the number of anomalies in the period of the increased geomagnetic activity when the accumulation windows of a few tens of days are used. The relative increase is roughly between 10 and 70 per cent. 

In the dataset recording the disturbances on transformers we see an increase, but we cannot reject the null hypothesis that the apparent increase is (probably) due to chance (the case of D8, high-voltage transformers), even though the probability is a bit lower than our rejection threshold. 

The merged dataset DX indicates an anomaly-rate increase in the high-activity periods consistently for all accumulation windows $W=30$~days or larger.  The interpretation of such a result is somewhat difficult because it contains anomalies registered on all sorts of devices and voltage levels. On the other hand, it may serve as a representative sample for the whole country covering 12 years of the electric power network performance.

These conclusive datasets also bear an indication that the increase is larger after the local maxima than before it, which is compatible with an interpretation that the increase indeed is due to the larger geomagnetic activity, where the effects of GICs show up with some delay. Such an interpretation is somewhat surprising for the datasets dealing with power lines (D7, D11, D12) or equipment connected to them (D10), where one does not expect cumulative effects due to the exposure to GICs, but rather an immediate response to the GICs entrance into the electric grid. 

Another hint that the disturbances indeed are caused by effects of geomagnetic activity would be if the disturbances occurring in networks of various operators would be somewhat correlated in time. Our code has an option to investigate such an issue. However, as evident from Fig.~\ref{fig:datasets}, the overlap period of all dataset is short, about three years. Such a length is way too short to obtain any statistically meaningful results. The poor common coverage is, unfortunately, a consequence of years-long delicate negotiations with the data providers. 

Our study is the first of this kind performed for the mid-latitude location. Many further studies are needed to fully understand our findings. For instance, the GIC modelling in the network together with an assessment of disturbances split to each of the power lines would give direct evidence about the influence of GICs on the equipment. The modelling of GICs is a goal of our ongoing study, our preliminary results indicate that there could be GICs as large as 50~A considered in the Czech power network. 

\acknowledgments
M.\v{S} was supported by the institute research project RVO:67985815. We are grateful to data providers for giving us an opportunity to exploit their logs of anomalies, namely to P.~Spurn\'y (\v{C}EPS), J.~Bro\v{z} and J.~Bu\v{r}i\v{c} (\v{C}EZ Distribuce), R.~Hanu\v{s} (PREdistribuce), and D.~Mezera and R.~B\'il\'y (E.ON Distribuce). The maintenance logs are considered strictly private by the power companies and are provided under non-disclosure agreements. The anonymised aggregate data series are available from \url{http://sirrah.troja.mff.cuni.cz/~svanda/2019SW002181/data.txt} with username \emph{spaceweather} and password \emph{2019SW002181}. The results presented in this paper rely on data collected at magnetic observatories. We thank the national institutes that support them and INTERMAGNET for promoting high standards of magnetic observatory practice. Magnetic data may be downloaded from www.intermagnet.org. We thank two anonymous referees for very useful comments that improved the message delivered by our paper. \emph{Author contributions}: M\v{S} designed the research which was then performed by TV within her MSc. project under the supervision of M\v{S}. Both authors contributed to the final manuscript.  

\bibliography{disturbances}

\end{document}